\documentclass[paper]{ieice}
\usepackage[pdftex]{graphicx,xcolor}
\usepackage[fleqn]{amsmath}
\usepackage{times}
\usepackage{algorithmic,algorithm}
\usepackage{color}

\field{}
\title[]{Learning-based WiFi Traffic Load Estimation in NR-U Systems}
\titlenote{This work was supported in part by the the National Natural Science Foundation of China under Grant No. 61771429, No. 61703368, and in part by Zhejiang University City College Scientific Research Foundation under Grant No. JZD18002. Zhejiang Lab's International Talent Fund for Young Professionals and Aviation Key Laboratory of Science and Technology on Fault Diagnosis and Health Management 20183333001 also provided support for the project.}
\authorlist{%
 \authorentry{Rui Yin}{m}{labelA}
 \MembershipNumber{2081027}
 \authorentry{Zhiqun Zou}{n}{labelB}
 \authorentry{Celimuge Wu}{m}{labelC}
 \MembershipNumber{1}
 \authorentry{Jiantao Yuan}{n}{labelD}
 \authorentry{Xianfu Chen}{n}{labelE}
 \authorentry{Guanding Yu}{n}{labelF}
}
\affiliate[labelA]
 {The author is with the School of Information and Electrical Engineering, Zhejiang University City College, Hangzhou, China. e-mail: yinrui@zju.edu.cn.}
\affiliate[labelB]
 {The author is with the School of Information and Electrical Engineering, Zhejiang University, Hangzhou, China.}
\affiliate[labelC]
 {The author is with Graduate School of Informatics and Engineering, The University of Electro-Communications, 1-5-1, Chofugaoka, Chofu-shi, Tokyo 182-8585, Japan.}
\affiliate[labelD]
 {The author is with the Institute of Ocean Sensing and Networking of the Ocean College, Zhejiang University, Zhoushan, China.}
\affiliate[labelE]
 {The author is with the VTT Technical Research Centre of Finland, Finland.}
 \affiliate[labelF]
 {The author is with the School of Information and Electrical Engineering, Zhejiang University, Hangzhou, China.}

\begin{document}
\maketitle

\begin{summary}
The unlicensed spectrum has been utilized to make up the shortage on frequency spectrum
in \emph{new radio} (NR) systems. To fully exploit the advantages brought by the unlicensed bands,
one of the key issues is to guarantee the fair coexistence with WiFi systems.
To reach this goal, timely and accurate estimation on the WiFi traffic loads is an important prerequisite.
In this paper, a \emph{machine learning} (ML) based method is proposed to detect the number of WiFi users on the unlicensed bands. An unsupervised \emph{Neural Network} (NN) structure is applied to filter the detected transmission collision probability on
the unlicensed spectrum, which enables the NR users to precisely rectify the measurement error and estimate the number of active WiFi
users. Moreover, NN is trained online and the related parameters and learning rate of NN are jointly optimized to estimate the number of WiFi users adaptively with high accuracy.
Simulation results demonstrate that compared with the conventional Kalman Filter based detection mechanism, the proposed approach has lower complexity and can achieve a more stable and accurate estimation.
\end{summary}

\section{Introduction}
The \emph{fifth generation} (5G) of mobile networks worldwide has already
provide low latency, high transmission rates and full coverage services to fulfill
the explosive diversified data demands, such as \emph{virtual reality} (VR), \emph{vehicles-to-everything} (V2X),
e-health, and smart grid. To further exploit the performance of 5G networks and expand the application scenarios of 5G transmission at close region,
one solution is to share the unlicensed
spectrum with WiFi system on the 5GHz \emph{unlicensed national information infrastructure} (U-NII)
band with approximately 600 MHz free bandwidth \cite{17_cui}.

Even though the unlicensed spectrum resource is free and juicy, it is for public use and mainly occupied by
WiFi networks nowadays. Therefore, the harmonious coexistence with WiFi networks must be guaranteed while providing service
on unlicensed bands in the cellular systems. The \emph{3rd generation partnership project} (3GPP)
has studied \emph{new Radio on Unlicensed bands} (NR-U) in Release 16 \cite{r16} after the standardization
of \emph{new Radio} (NR) in 5G. Unlike the conventional centralized \emph{medium access control} (MAC)
protocol employed in the NR on licensed bands, NR-U has to adopt the contention based MAC to ensure the
fair sharing on unlicensed spectrum.

Before the start of the study items on NR-U, most of works have concentrated on the channel access scheme
design under the constraint of fair coexistence with WiFi in the \emph{long term evolution on unlicensed spectrum} (LTE-U).
The \emph{licensed-assisted access} (LAA) proposed in Release 13 has been considered as the main candidate for the
LTE-U networks \cite{r13}. In \cite{16_yin}, an adaptive back-off scheme has been proposed for the LAA based
small cell networks to avoid the access
collision with WiFi network under the assumption that the WiFi traffic load is known in advance.
In \cite{18_cha}, a deep reinforcement learning algorithm based on
long short-term memory cells has been applied to learn the cellular traffic loads and the utilization
pattern of the unlicensed channels by the WiFi networks.
Therefore, the LAA small \emph{base station} (BS) can decide how to use the unlicensed channels accordingly.
The application of LAA in NR-U networks has been investigated in
\cite{19_song}, where the cooperation among small cells is leveraged
to decrease the collision probability and increase the system capacity.
Moreover, the edge computing technologies are applied to integrate both licensed and unlicensed spectrum resources in \cite{19_wu}, where a joint aggregation, caching, and decentralization scheme has been proposed to enhance system performance.

The obtainable accurate WiFi traffic load at the
NR cellular systems is a key to guarantee the fair coexistence with WiFi networks while improving
the \emph{spectrum efficiency} (SE) on unlicensed spectrum \cite{16_Ko,17_Chen}. However, most of existing literatures configure access parameter to ensure fairness on the assumption that the WiFi traffic load is known \cite{18_Mag,19_Liu,19_Hu,18_Yuan,20_Zhao}.
As the WiFi traffic load is mainly decided by the active WiFi users, only several works study how to estimate the real traffic condition.
For instance, an extended \emph{Kalman filter} (KF) scheme has been applied to estimate the
number of active WiFi users based on  the measured collision probability on the
corresponding unlicensed channels in \cite{03_Bianchi}. Results show that it works well when the WiFi traffic load is low, however, the estimation error can not be ignored when the traffic load is heavy. Moreover, its high computational complexity requires more computing resources and limits its application in practice.
Some improved KF based algorithms are proposed to predict the number of WiFi terminals and \emph{signal noise ratio} (SNR) \cite{13_Qin,10_Zheng,10_Haider,16_chen}, where WiFi transmission model is further improved. However, the KF based methods perform well only on linear, discrete and finite-dimension states and the problem mentioned above still exists. Therefore, a more effective algorithm needs to be designed to ensure the accuracy of the estimation in a dynamically loaded WiFi environment.

In this paper, to avoid high computational complexity and improve the estimation accuracy, an online trained \emph{neural network} (NN) structure is proposed to learn the WiFi traffic load information. With the excellent extensibility of NN, it can estimate and track the changes on the number of WiFi users faster as well as accurately under various WiFi traffic load.
With the estimation of WiFi traffic load, the proposed algorithm can maximize the spectrum efficiency of the unlicensed spectrum while ensuring the best performance of WiFi system.

The main contributions of the paper are summarized as follows:
\begin{enumerate}
\item[1.] A targeted NN with simple three hidden layers is designed to estimate the number of WiFi users. Attribute to
  its simple structure, the sensitivity of the system is guaranteed and the computational complexity can be reduced significantly.
\item[2.] The effective threshold mechanism based on \emph{cumulative summary} (CUSUM) is applied to track the change on the number of WiFi users by the proposed scheme. Moreover, an adaptive learning rate and loss function adjustment are proposed to accelerate the convergence speed of the NN as well as ensure the stability of the output.
\item[3.] Numerical results are provided to verify the effectiveness of the NN based learning scheme on the estimation accuracy and convergence speed.
\end{enumerate}

The rest of this paper is organized as follows. We introduce the system model in Section~\ref{s2}. The extended KF based method is introduced in Section~\ref{s3plus}. Section~\ref{s3} describes the proposed NN based model. The simulation results are analyzed in Section~\ref{s4} and conclusions are arranged in Section~\ref{s5}.

\section{System Model}\label{s2}
\begin{figure}[!t] \centering
\begin{centering}
\includegraphics[width=0.45\textwidth]{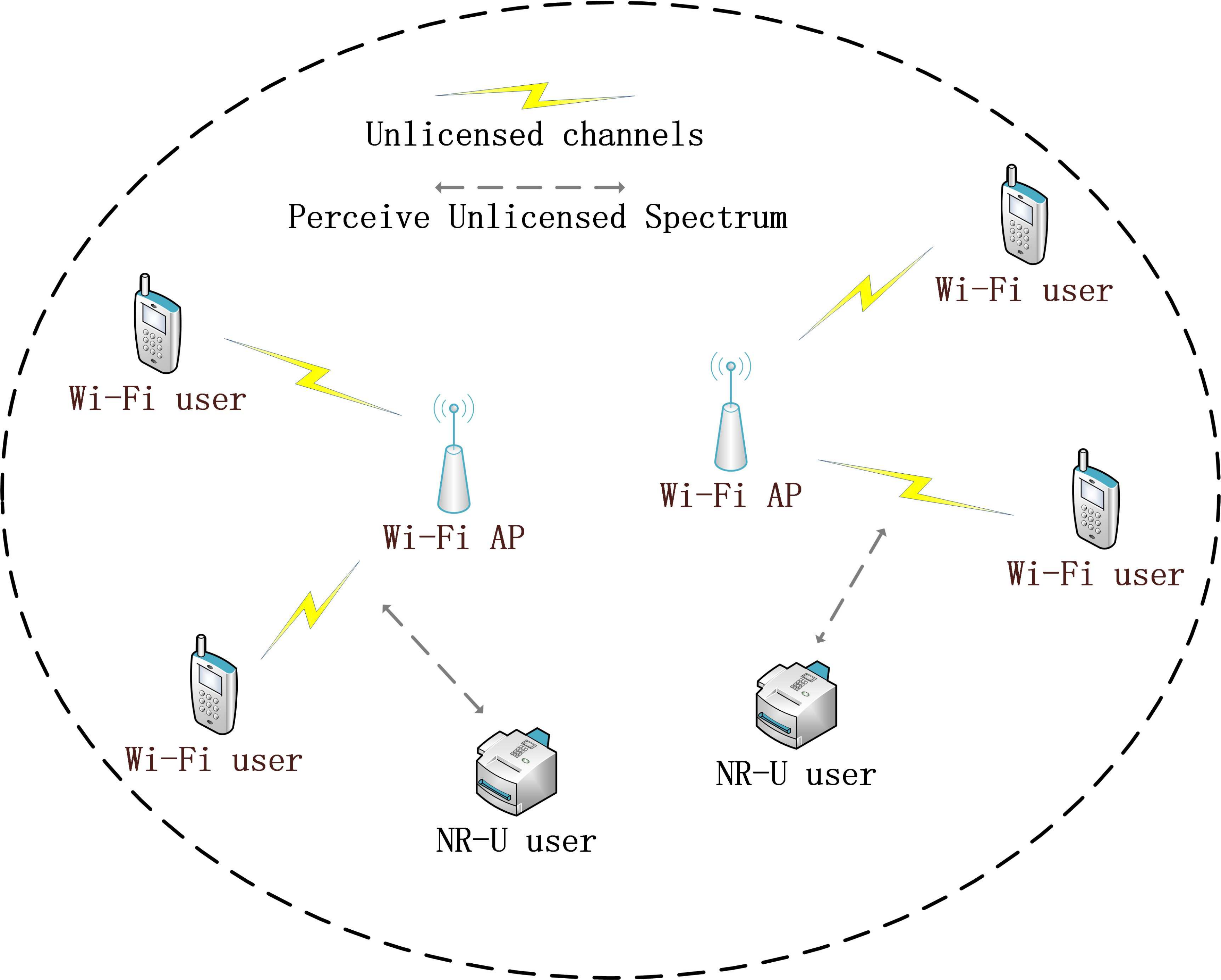}
\vspace{0.2cm}
\caption {System model, the NR-U users sense the unlicensed channels before using them.}\label{systemmodel}
\end{centering}
\end{figure}

In the paper, we assume that there are multiple WiFi \emph{access points} (APs) with the associated WiFi users transmitting on the same unlicensed channel and a set of NR users detect and sense the unlicensed channel before using it, as shown in Fig~\ref{systemmodel}.
In addition, both the WiFi APs and WiFi users adopt the \emph{channel sensing multiple access with collision avoidance} (CSMA/CA) based \emph{distributed coordination function} (DCF) to access unlicensed spectrum.
To model the system mathematically, WiFi APs that use the same unlicensed channel are represented by set $\mathcal{W}=\{w_0,w_1,...,w_{N-1}\}$, where $N$ is the number of WiFi APs under the coverage of NR-U system. All WiFi APs in $\mathcal{W}$ employ IEEE 802.11n protocol.
Define a set as $\mathcal{N}=\{n_0,n_1,...,n_{N-1}\}$ to demonstrate the number of WiFi users associated with the corresponding WiFi APs, where $n_k$ represents the number of WiFi users served by WiFi AP $w_k$. Moreover, $\mathcal{U}$ is used to represent the NR cellular users who
need to estimate the number of WiFi users on the unlicensed channel. For all users in $\mathcal{U}$, the detection process is independent, so we focus on a single NR user defined as $u$ hereinafter.

\subsection{Number of Competing WiFi Users}
According to the IEEE 802.11n protocol, WiFi system adopts binary slotted exponential back-off scheme to minimize the collision probability on unlicensed channel.
In \cite{00_Bianchi}, Markov chain models have been employed to
derive the closed-form expression on the collision probability, transmission probability and the
system throughput.
Based on Markov chain models,
the probability that a WiFi user transmits a data packet at any time slot is given by
\begin{eqnarray}\label{eq1}
\small
\tau = \frac{2(1-2P)}{(1-2P)(G+1)+PG(1-(2P)^{m})},
\end{eqnarray}
where $G$ and $m$ are the size of back-off contention window and the maximum back-off contention stage, respectively. These two parameters are fixed in the physical layer according to IEEE 802.11n. $P$ is the transmission collision probability experienced by each WiFi user, which can be calculated as
\begin{eqnarray}\label{eq2}
\small
P = 1 - (1 - \tau)^{n-1},
\end{eqnarray}
where $n$ is the number of WiFi users competing for the same unlicensed channel,
which is equal to
{\small{$n = \sum_{k=0}^{N-1}n_k$}}.
By substituting \eqref{eq1} into \eqref{eq2}, we can write $n$ as a function of $P$, which is
\begin{eqnarray}\label{eq4}
\small
\lefteqn{n = f(P)}\quad \nonumber\\
 &=& 1 + \frac{\log(1-P)}{\log(1-\frac{2(1-2P)}{(1-2P)(G+1)+PG(1-(2P)^{m})})}.
\end{eqnarray}
Since $G$ and $m$ are limited and configurable by the specification in IEEE 802.11n, equation \eqref{eq4} can be utilized at NR user $u$ to predict the value of $n$ based on the observation of the transmission collision probability on the corresponding unlicensed channel.
The key factor affecting the accuracy on the prediction of $n$ is the obtainable value of $P$.

\subsection{Transmission Collision Probability}
According to \eqref{eq4}, if the value of $P$ can be obtained precisely, the estimation on $n$ may achieve
high accuracy.
In this regard, NR user $u$ first needs to measure the value of $P$ by sensing the unlicensed channel during the WiFi transmission.
At time slot $t$, $u$ records the number of busy sub-frames of WiFi transmission on the unlicensed channel as $K_{b,t}$, and collision sub-frames, $K_{c,t}$. Let $K_{a}$ denote the number of all the observed sub-frames during the whole time slot. Then, the measurement result of $P$ at time slot $t$ is given by
\vspace{0.25cm}
\begin{eqnarray}\label{eq5}
\small
\hat P_t = \frac{K_{b,t}+K_{c,t}}{K_{a}}.
\end{eqnarray}
Define $\hat n_t$ as the measurement result of the number of WiFi users, which can be derived from \eqref{eq4} with $\hat P_t$, denoted as $\hat n_t = f(\hat P_t)$. Apparently, due to the measurement error on $P$ in \eqref{eq5}, Only a few estimates are unable to reflect the real number of WiFi users accurately. Moreover, according to \eqref{eq4}, a tiny change on $P$ may cause huge fluctuation on the estimation of $n$.
Therefore, the accuracy on the estimation of $P$ is crucial to obtain $n$ precisely.
In order to guarantee the performance, the conventional method to detect $n$ is mainly based on the extended \emph{Kalman Filter} (KF), where KF is utilized to revise output according to the continuous observation of $\hat P_t$ to obtain accurate estimation of $n$.
However, there are a large number of circulation loop calculations in the KF based algorithm, which leads to high computational complexity. Another weakness is that the extended KF can only achieve low accuracy when the number of WiFi users is large.
The algorithm and performance of KF based method will be further analyzed in the next section.

\section{Extended Kalman Filter based Mechanism}\label{s3plus}
In order to obtain the number of WiFi users based on the transmission collision probability measured in \eqref{eq5}, the extended KF is the most commonly used mechanism.
Therein, the number of WiFi users, {\small{$n = \sum_{k=0}^{N-1}n_k$}}, in the system is treated as the state variable of KF and the
collision probability calculated in \eqref{eq5} is the measurement result used to refine the estimation on $n$.
The state estimation of the framework as well as the output of KF at time slot $t$ is expressed as $n^{k}_{t}$,
which means the estimation of $n$ by the extended KF.
Let $V_{t}$ be the estimation error variance of $n^{k}_{t}$ and the updating formulas of KF are given by
\begin{eqnarray}\label{eq3_1}
n^{k}_{t} = n^{k}_{t-1} + K_{t}Z_{t},
\end{eqnarray}
and
\begin{eqnarray}\label{eq3_2}
V_{t} = (1 - K_th'(n^{k}_{t-1}))(V_{t-1}+Q),
\end{eqnarray}
herein $Z_t$ reflects the error between the estimation result and the measured value of $P$, which is computed by
\begin{eqnarray}\label{eq3_3}
\hat Z_{t} = \hat P_{t} - h(n^{k}_{t-1}),
\end{eqnarray}
where $\hat P_{t}$ is the measured value that $u$ measures based on \eqref{eq5} at time slot $t$ and $h(\cdot)$ is the anti-function of $f(\cdot)$ in \eqref{eq4}, $h(n^{k}_{t-1})$ represents the estimation value of collision probability according to the number of WiFi users predicted at the last time slot. $K_{t}$ is named as Kalman gain and is calculated as
\vspace{0.25cm}
\begin{eqnarray}\label{eq3_4}
K_{t} = \frac{h'(n^{k}_{t-1})(V_{t-1}+Q)}{(h'(n^{k}_{t-1}))^2(V_{t-1}+Q)+R_t},
\end{eqnarray}
where $Q$ represents the system noise estimated by KF and it is set according to the real case scenario.
When the value of $Q$ is large, the extended KF is sensitive to the variation of the measured results. On the other hand,
when the value of $Q$ is small, system is not sensitive to the noise and tends to output stable values.
$h'(\cdot)$ is the derivative function of $h(\cdot)$, which reflects the sensitivity of
actual measurement. Besides, $R_t$ denotes the variance of the measurement on $P$ , which is given by
\vspace{0.25cm}
\begin{eqnarray}\label{eq3_5}
R_{t} = \frac{(1-h(n^{k}_{t-1}))h(n^{k}_{t-1})}{K_{a}}.
\end{eqnarray}
\vspace{0.25cm}

As mentioned above, the value of $Q$ has obvious impact on the prediction results of the extended KF.
When the number of WiFi users is stable, $Q$ requires to be set small, which can provide accurate and
robust prediction. On the contrary, when the number of WiFi users changes,
a large $Q$ value should be chosen to improve the sensitivity of the system.
In order to realize the adaptive adjustment of $Q$, \emph{cumulative summary} (CUSUM) based threshold method is utilized, where $Q$ is set to $Q^+$ when the threshold is triggered and otherwise set to $Q^-$.

\begin{figure}[!t] \centering
\begin{centering}
\includegraphics[width=0.45\textwidth]{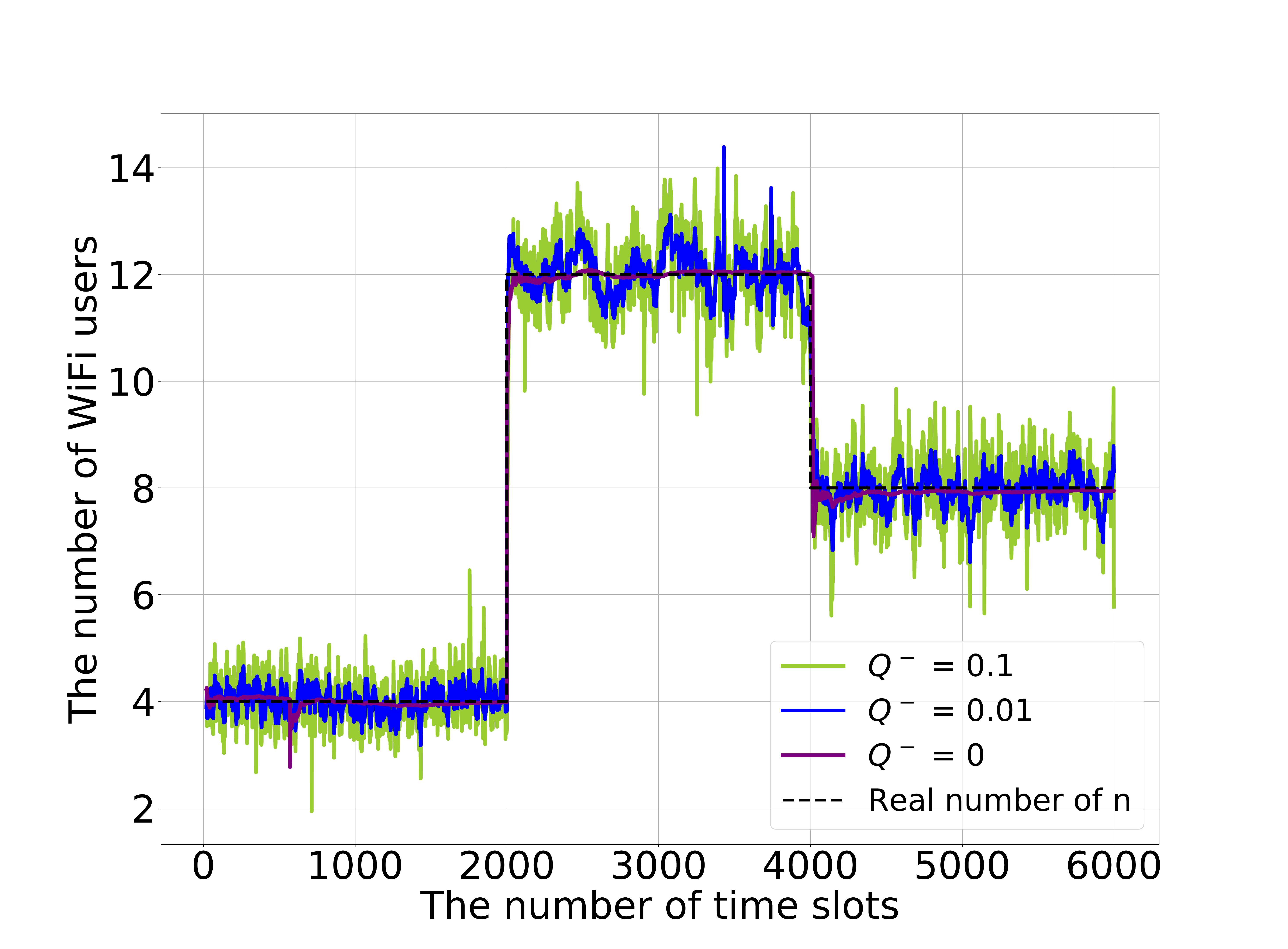}
\caption {The estimation results by the extended KF scheme when $n$ is smaller than 12.}\label{Kalmanresult1}
\end{centering}
\end{figure}

\begin{figure}[!t] \centering
\begin{centering}
\includegraphics[width=0.45\textwidth]{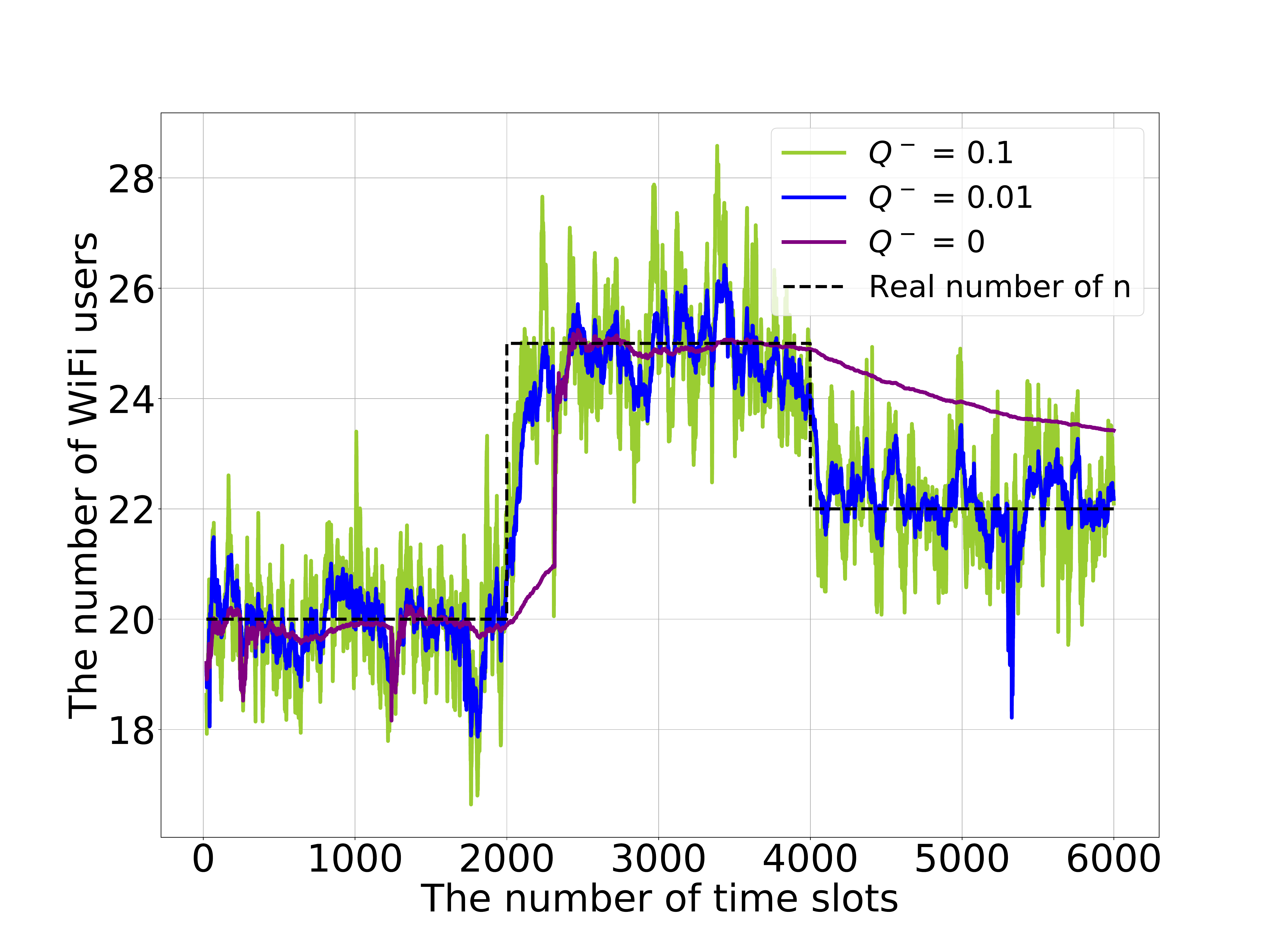}
\caption {The simulation results by the extended KF scheme when $n$ is larger than 20.}\label{Kalmanresult2}
\end{centering}
\end{figure}

The simulation results of the estimation performance by the extended KF based scheme are demonstrated in Fig.~\ref{Kalmanresult1} and Fig.~\ref{Kalmanresult2}. The back-off contention window size $G$ and the maximum back-off contention stage $m$ of WiFi APs are set to $32$ and $3$, respectively.
Furthermore, $Q^+$ is fixed at $4$ and performances with different $Q^-$ values are compared in the simulation results. During the simulation, the number of WiFi users is changed in every 2000 time slots.

\begin{figure}[!t] \centering
\begin{centering}
\includegraphics[width=0.45\textwidth]{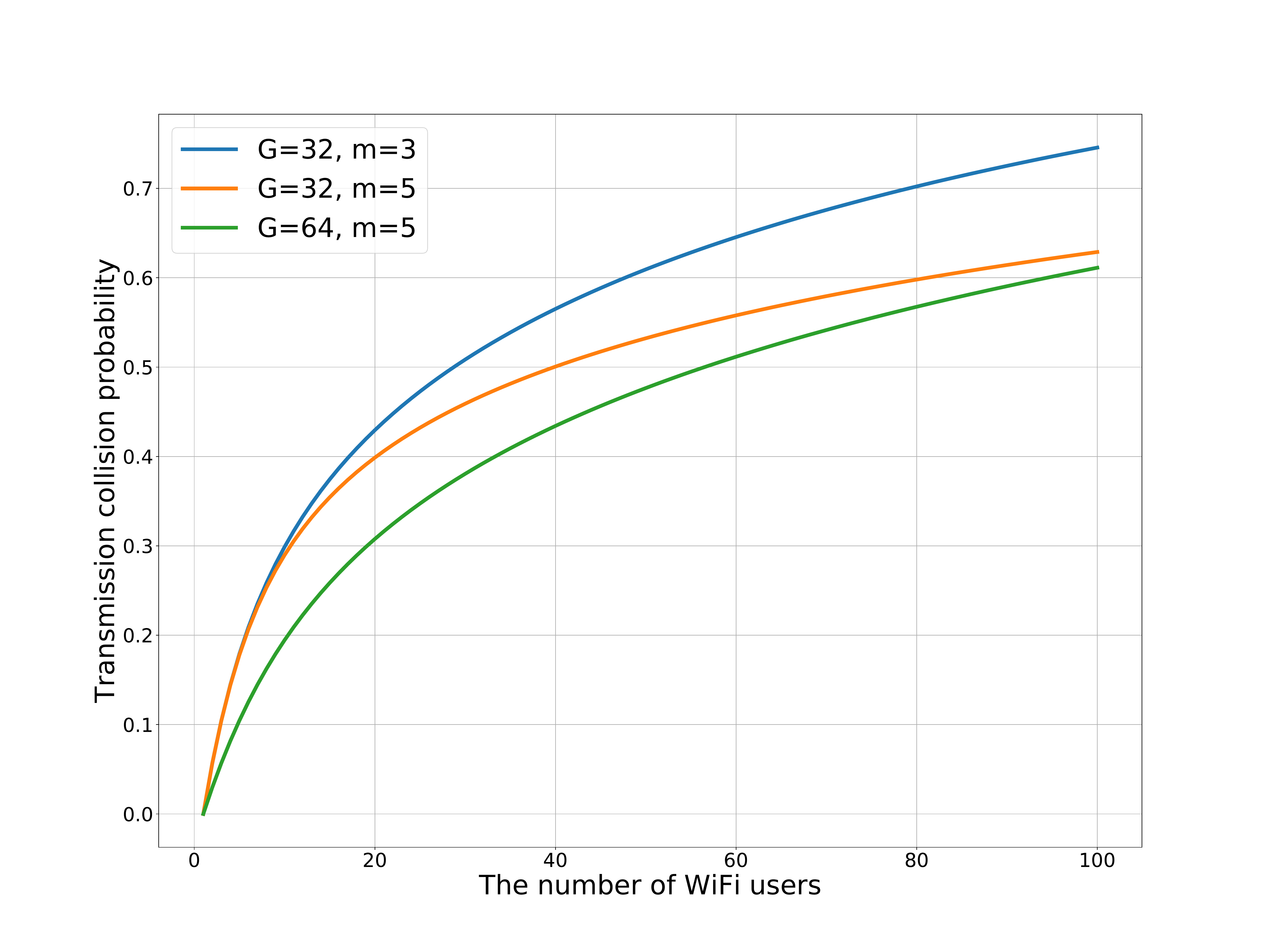}
\caption {The relation diagram of $P$ and $n$ in WiFi system.}\label{wifiPN}
\end{centering}
\end{figure}

According to the performance illustrated in Fig.~\ref{Kalmanresult1}, we can observe that the extended KF based method can achieve accurate estimation when the number of WiFi users, $n$, is small. In addition, it can be found that the smaller $Q^-$ value can provide more stable output.
However, as the number of active WiFi users increases,the estimation value becomes unreliable and fluctuates significantly, as shown in Fig.~\ref{Kalmanresult2}. The reason is that when the number of WiFi users is large, the small difference on $n$ will cause tiny variation of $P$, as shown in Fig.~\ref{wifiPN}, and conversely, the noise generated when measuring $\hat P_{t}$ will be amplified to affect the estimation on $n$. As a result, the huge measurement error prevents the Kalman filter from outputting smoothly and accurately.

In addition, as mentioned above, $h(\cdot)$ is the anti-function of \eqref{eq4} and $h'(\cdot)$ is the derivative function of $h(\cdot)$. However, we are unable to derive the close-form expression of both $h(\cdot)$ and $h'(\cdot)$ from \eqref{eq4}.
The value of $h(n^{k}_{t-1})$ and $h'(n^{k}_{t-1})$ can only be computed by numerical methods. Therefore, a large number of circulation loop calculations need to be executed in order to achieve the accurate results, which leads to high computational complexity as well as weak real-time ability. The detailed time consumption to run the extended KF scheme will be analyzed in Section~\ref{s4}.

In the next section, in order to improve the prediction accuracy as well as decrease the computational complexity,
an online unsupervised training neural network is utilized to estimate the number of active WiFi users.
Particularly, to realize more steady and adaptive detection, a specific loss function is deployed to train the neural network.

\section{Learning based WiFi Users Number Estimation}\label{s3}
\begin{figure}[!t] \centering
\begin{centering}
\includegraphics[width=0.45\textwidth]{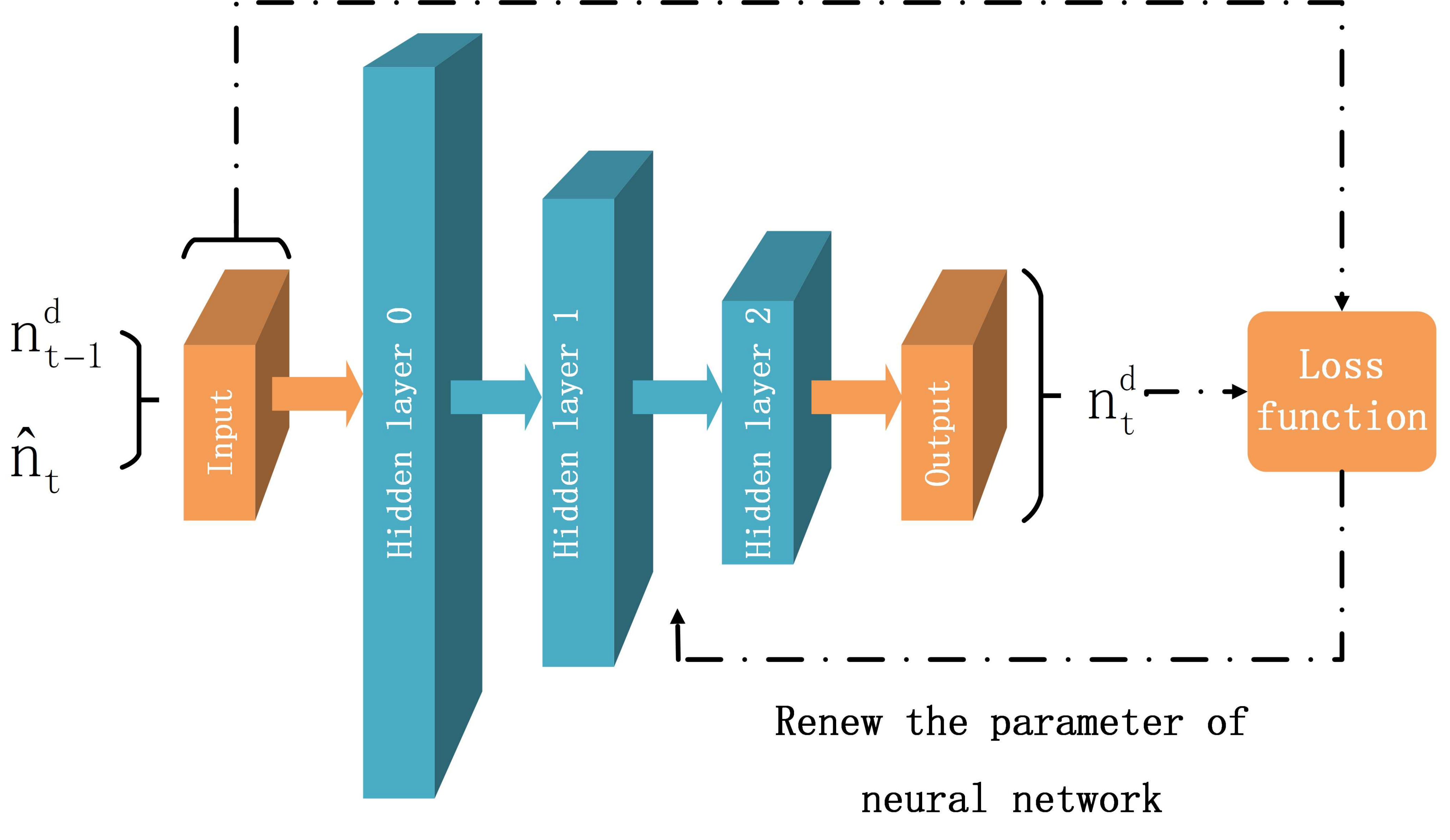}
\caption {The structure of NN based network.}\label{neuralnetwork}
\end{centering}
\vspace{-0.5cm}
\end{figure}

To reduce the computational complexity and track the change on
the number of WiFi users adaptively, a \emph{machine learning} (ML) based approach is developed
in this section. Due to the excellent robustness and the low complexity of NN, an online unsupervised learning NN architecture is deployed to reduce the computational complexity as well as pursue high prediction and tracking accuracy in a congested WiFi system. The structure of the proposed NN is demonstrated in Fig~\ref{neuralnetwork}.

The duration that user $u$ evaluates $P$ based on \eqref{eq5} and outputs the prediction of $n$ is denoted as a decision time slot
in the proposed approach. Then, at time slot $t$, the output of the forward propagation for NN can be expressed as
\begin{eqnarray}\label{eq13}
\small
\mathbf o_t = F([\mathbf i_t],\theta_{t-1}) = n_{t}^d,
\end{eqnarray}
where $\mathbf i_t=[n_{t-1}^d,\hat n_{t}]$ is the input to the NN.
$n_{t}^d$ is the estimation of $n$ by NN at $t$, $\theta_t$ represents all the weight and bias parameters in NN and is updated based on the gradient descent algorithm, which is given by
\begin{eqnarray}\label{eq14}
\small
\theta_t = \theta_{t-1} - l\frac{\partial L_t}{\partial \theta_{t-1}},
\end{eqnarray}
where $l$ denotes the learning rate, $L_t$ is the corresponding loss function as well as the objective function of NN. Since conventional supervised learning based approach require large amounts of labeled data for training and has poor adaptability in a variable environment, we propose an adaptive objective function to train NN unsupervised in a changing system, here the objective function $L_t$ is defined as
\begin{eqnarray}\label{eq15}
\small
L_t = \frac{\alpha (n_{t}^d - \hat n_{t})^2}{2} + \frac{\beta (n_{t}^d - n_{t-1}^d)^2}{2}.
\end{eqnarray}
The first item in the right-hand-side of \eqref{eq15} represents the measurement error while the second item denotes the prediction error.
Parameters $\alpha$ and $\beta$ are the weight factors associated with the measurement error and prediction error, respectively. When $\alpha$ is set to a small value while $\beta$ is assigned with a large value, the scale of the loss value is mainly decided by the second item of \eqref{eq15}. As a consequence, the output of NN tends to converge to the historical prediction value, which implies that NN is not sensitive to the change of $n$, but tends to be stable. On the other hand, when $\alpha$ is set to a large value while $\beta$ is small, the measurement value $\hat n_{t}$ has a larger contribution on the loss function and the output of NN will be trained towards $\hat n_{t}$, which implies that NN is sensitive to the change of the real measurement result.
Therefore, $L_t$ can achieve a good tradeoff between the estimation in the previous time slot and current measurement on $n$.

Apparently, if the number of active WiFi users, $n$, does not change, the expectation value of $\hat n_{t}$ and $n_{t}^d$ are both equal to $n$, i.e. $\mathbf E\{\hat n_{t}\}=\mathbf E\{n_{t}^d\}=n$.
However, the number of active WiFi users changes over time, which causes that the expectation of $n_{t}^d$ deviates from $n$.
To ensure stable estimation as well as sensitive adjustment of the NN, objective function should be changed accordingly.
If the WiFi traffic is stable, the mean value of the loss is equal to $0$, i.e. $\mathbf E\{L_t\}=0$. Therefore, to achieve a small $L_t$, the value of $L_t$ should be mainly decided by the historical prediction value and the influence of the measurement results should be scaled down.
On the contrary, when the WiFi traffic load changes, the mean value of loss function will not be $0$ until NN converges to a new stable state. Therefore, during the convergence stage, the value of $L_t$ has to concentrate more on the measurement.
In consequence, parameters, $\alpha$ and $\beta$, should change adaptively according to the variation on the WiFi traffic load.
Herein, the threshold mechanism based on CUSUM is also employed to estimate the stability condition of the system and adjust the parameters, where the threshold parameter, $g_t^d$, is defined as
\begin{eqnarray}\label{eq16}
\small
g_t^d =
\begin{cases}
\max(0,g_{t-1}^d+L_t-q), & \mbox{if }g_{t-1}^d \le e^d,\\
0+L_t-q, & \mbox{otherwise}.
\end{cases}
\end{eqnarray}
Herein, $g_t^d$ is initialized to $0$ when $t = 0$ and $q$ is a constant value which denotes the tolerance to the estimation error, $e^d$ is the constant trigger value that determines whether the state of the system has changed. Based on this mechanism, if $g_t^d > e^d$, then $\alpha$ will be set to be a large value $\alpha^+$ and $\beta$ is adjusted to a small value $\beta^-$. Instead, if $g_t^d \le e^d$, $\alpha$ will turn to a small value $\alpha^-$ and $\beta$ will be changed to $\beta^+$.
Moreover, in order to better apply this adaptive mechanism as well as speed up the convergence process, the learning rate,
$l$ of the NN, is also adjusted when the change on the number of active WiFi users
is detected, which is given by
\begin{eqnarray}\label{eq17}
\small
l =
\begin{cases}
l^-, & \mbox{if }g_t^d \le e^d,\\
l^+, & \mbox{otherwise}.
\end{cases}
\end{eqnarray}
In the above learning rate updating mechanism, when $g_t^d$ meets the trigger value $e^d$, denoting that $u$ has detected the change on the number of active WiFi users, the loss value computed by \eqref{eq15} is amplified and the learning rate, $l$, will
increase accordingly to train the network fast and efficiently. On the contrary, if $g_t^d \le e^d$, the NN regards that the number of
active WiFi users does not change, therefore, both $\alpha$ and $l$ are set to small values.
In consequence, the output of NN tends to be stable and fine-tuned when WiFi traffic load is stable.

Besides, to further stabilize the NN output when WiFi traffic load is steady, a specific fully-connected structure of NN is designed, where the number of neurons decreases with the number of layers. In such a structure, most weight and bias parameters lie in the lower layers of the NN. Due to the chain rule of the back propagation training approach, the gradient of the bottom parameters is smaller. Therefore, during
the training process, most parameters will be merely trained and the whole NN can basically keep steady.

The complete learning process via the NN is concluded in Algorithm~\ref{nn_rt}, when $t=0$, $\mathbf i_0$ is set to $[0, \hat n_{0}]$.
$T$ is the entire time period for $u$ to estimate $n$. Here a time slot includes the time that $u$ detects $\hat P_t$ and the time to execute the algorithm.
At time slot $t$, the forward propagation on NN is first executed to get the output $\mathbf o_t = n_t^d$. Then $L_t$ and $g_t^d$ are calculated according to \eqref{eq15} and \eqref{eq16}, respectively. Based on $g_t^d$, NN related parameters, $\alpha$, $\beta$, and learning rate $l$, are set to the corresponding values. Finally, loss value is updated and the back propagation algorithm is implemented to train the NN.

\begin{algorithm}[!t]\label{a1}
\small{
\caption{NN-based WiFi traffic load estimation algorithm}\label{nn_rt}\centering
\begin{algorithmic}[1]
\STATE Initialize the structure and the relevant parameters of NN;
\FOR{$t \in T$}
\STATE The input of NN is $\mathbf i_t = [n_{t-1}^d, \hat n_{t}]$;
\STATE The forward propagation of neural network is carried out to obtain the output $\mathbf o_t = n_t^d$;
\STATE Calculate the loss value on the basis of \eqref{eq15};
\STATE Renew the value of $g_t$ based on \eqref{eq16};
\IF{$g_t^n > h$}
\STATE $l = l^+$, $\alpha = \alpha^+$, $\beta = \beta^-$;
\ELSE
\STATE $l = l^-$, $\alpha = \alpha^-$, $\beta = \beta^+$;
\ENDIF
\STATE Calculate the loss value on the basis of \eqref{eq15};
\STATE Train the NN based on gradient descent algorithm;
\ENDFOR
\end{algorithmic}}
\end{algorithm}

\begin{table}\centering
\small{
\caption{Simulation parameters}\label{sim_para}
\vspace{0.1cm}
\begin{tabular}{|c|c|c|}
  \hline
  Parameters & Value\\
  \hline
  Measurement weight $\alpha^+/\alpha^-$ & $0.99/0.01$\\
  \hline
  Experience weight $\beta^+/\beta^-$ & $0.99/0.01$\\
  \hline
  Learning rate $l^+/l^-$ & $0.1/0.01$\\
  \hline
  Threshold activation value $e^d$ & $20$\\
  \hline
  Tolerance parameter $q$ & $0.1$\\
  \hline
\end{tabular}}
\end{table}

\section{Simulation Results} \label{s4}
In this section, the performance of the proposed scheme is verified by the numerical simulations.
In the simulation setup, the parameters related to the proposed NN in Section III is given in Table~\ref{sim_para}.
To guarantee the high sensitivity and low complexity, a four-layer fully connected NN structure is employed and the related parameters are shown in
Table~\ref{NN_para}.

\begin{table}\centering
\small{
\caption{NN parameters}\label{NN_para}
\vspace{0.1cm}
\begin{tabular}{|c|c|c|}
  \hline
  Parameters & Value\\
  \hline
  Number of neurons & $32/16/8/4$\\
  \hline
  Active function & $\tanh/\tanh/\tanh/none$\\
  \hline
  Optimizer & Adam Optimizer\\
  \hline
\end{tabular}}
\end{table}

The parameters $G$ and $m$ of WiFi APs are also set to $32$ and $3$, respectively.
The number of active WiFi users changes in every $2000$ time slots.

\begin{figure}[!t] \centering
\begin{centering}
\includegraphics[width=0.45\textwidth]{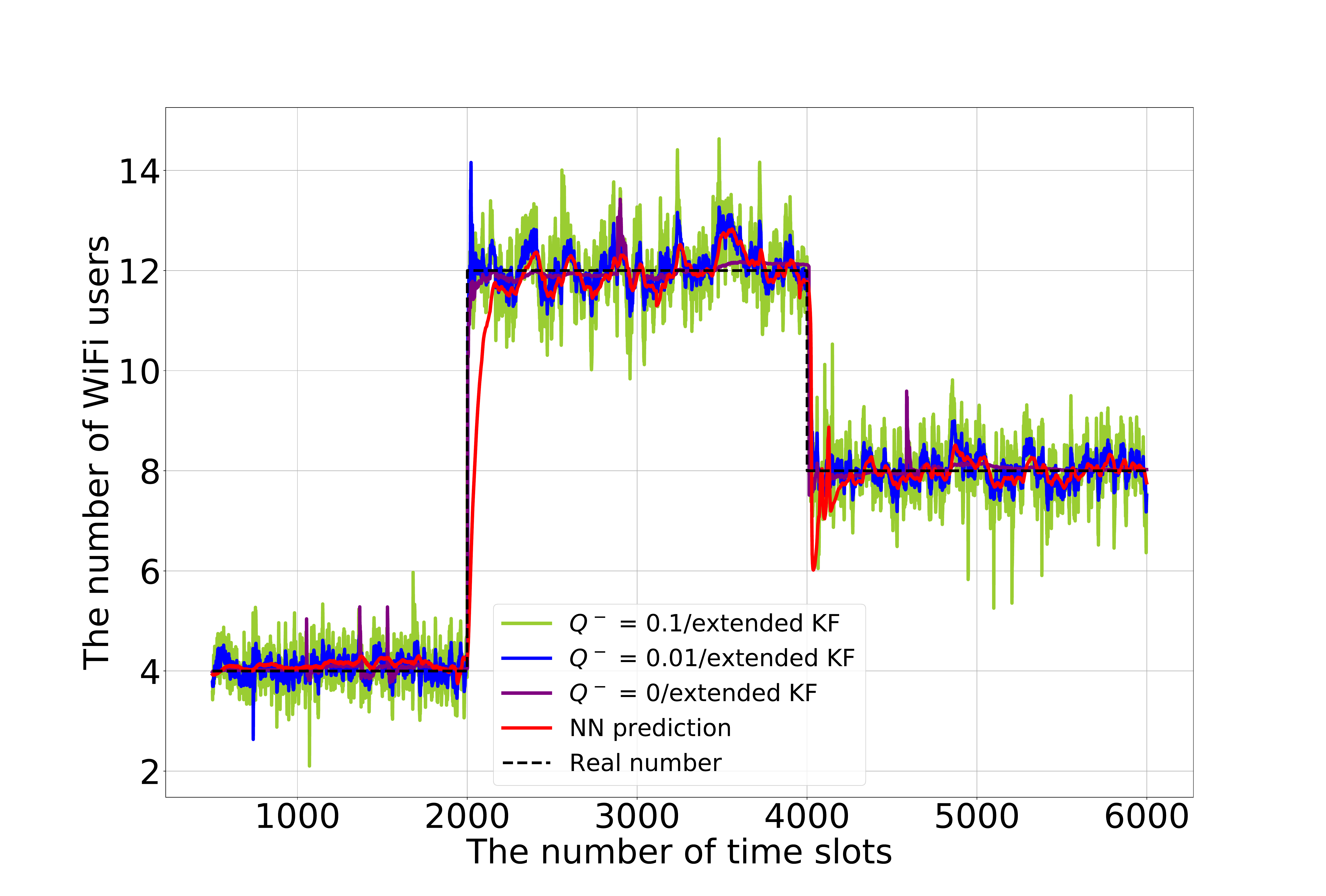}
\caption {Simulation result when $n$ is smaller than 12.}\label{nnsimulation1}
\end{centering}
\vspace{-0.5cm}
\end{figure}

\begin{figure}[!t] \centering
\begin{centering}
\includegraphics[width=0.45\textwidth]{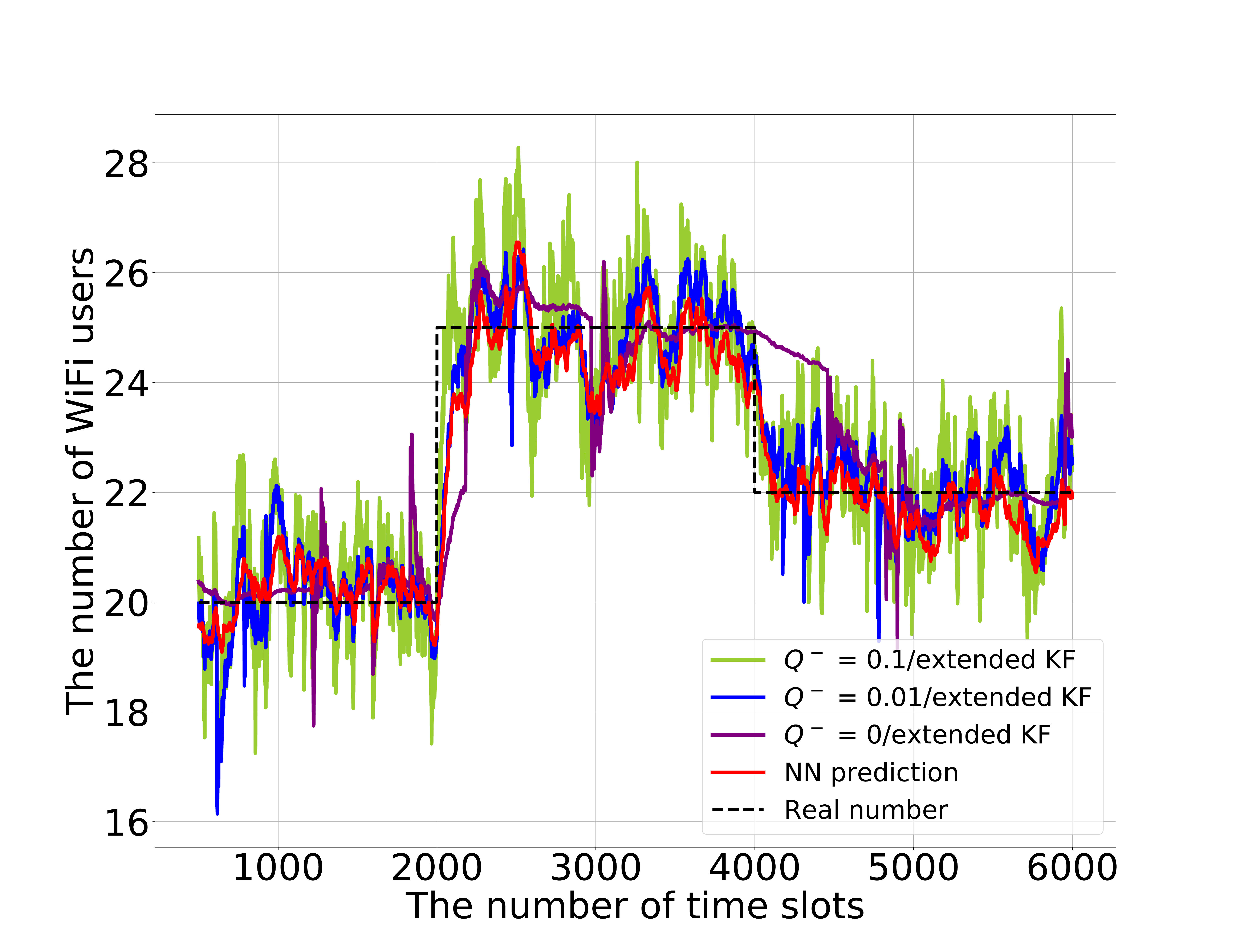}
\caption {Simulation result when $n$ is larger than 20.}\label{nnsimulation2}
\end{centering}
\vspace{-0.5cm}
\end{figure}

The estimation on the WiFi traffic load by Algorithm 1
and the extended KF based method is illustrated in Fig.~\ref{nnsimulation1} and Fig.~\ref{nnsimulation2}, respectively.
In Fig.~\ref{nnsimulation1} where the number of active WiFi users is smaller than $12$.
We can observe that the extended KF scheme with $Q^- = 0$ provides the most steady and accurate output and the performance of NN is
very close to that of the extended KF scheme.
On the other hand, when the number of WiFi users is larger than $20$, the estimation accuracy by the proposed
approach is much higher than the extended KF methods, as shown in Fig.~\ref{nnsimulation2}.
As analysed in Section~\ref{s3plus}, when working with a large number of active WiFi users (over 20 in our system), the variation of $n$ will cause less change on $P$ and the NN in the proposed scheme is able to better detect this subtle change than the extended KF scheme. Although $Q^- = 0$ remains the steady output of the extended KF, it could hardly response to the variation in time.
As for convergence, based on the simulation results, we can find that the output of the network is basically fine-tuned, which means the loss value of NN is very small in most of the time. The loss value only becomes larger after the error has accumulated to be higher than the threshold and then the output of NN will converge to the test value.

\begin{figure}[!t] \centering
\begin{centering}
\includegraphics[width=0.45\textwidth]{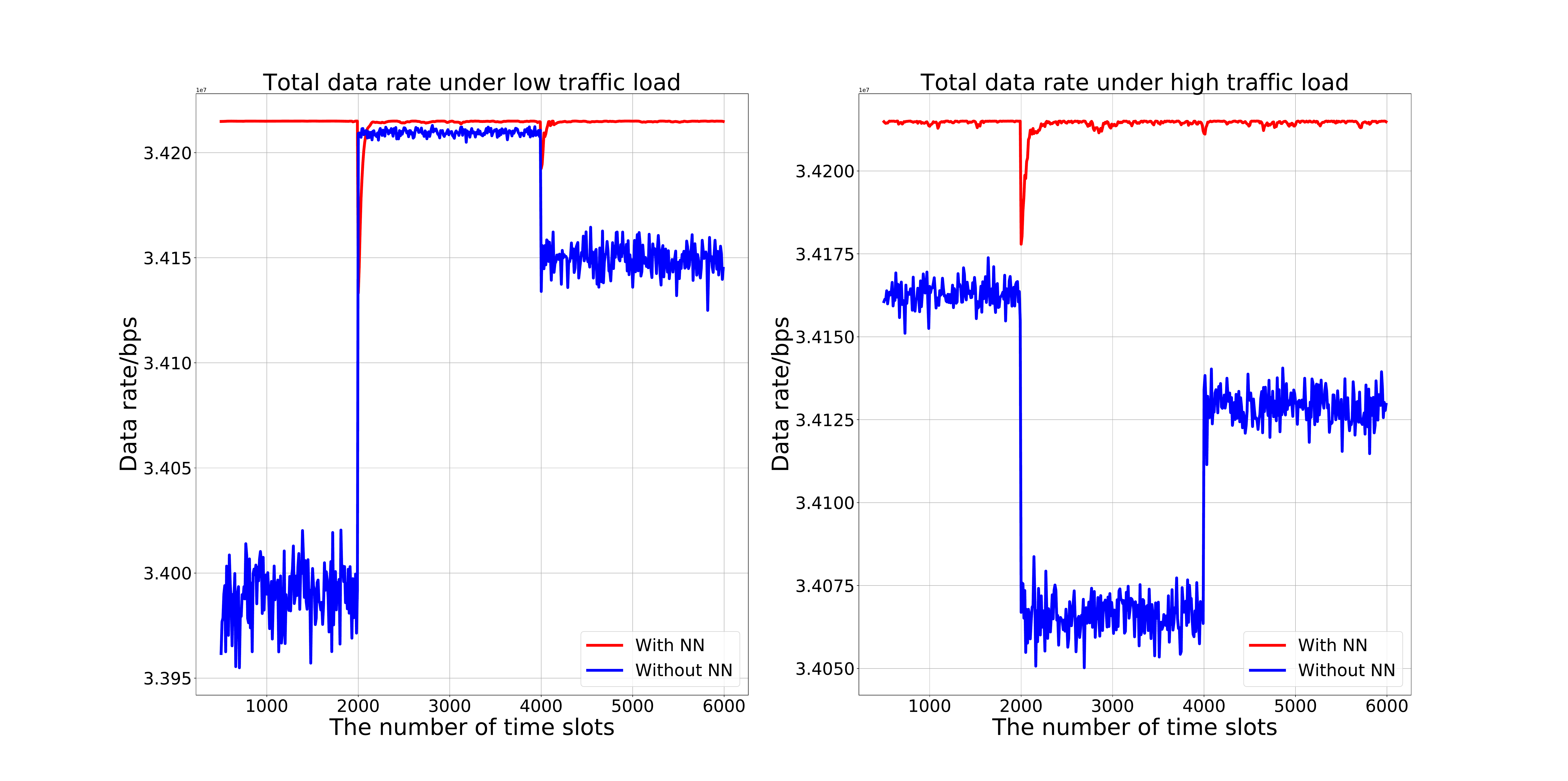}
\caption {Total data rates in the unlicensed channel.}\label{totalDR}
\end{centering}
\end{figure}

\begin{figure}[!t] \centering
\begin{centering}
\includegraphics[width=0.45\textwidth]{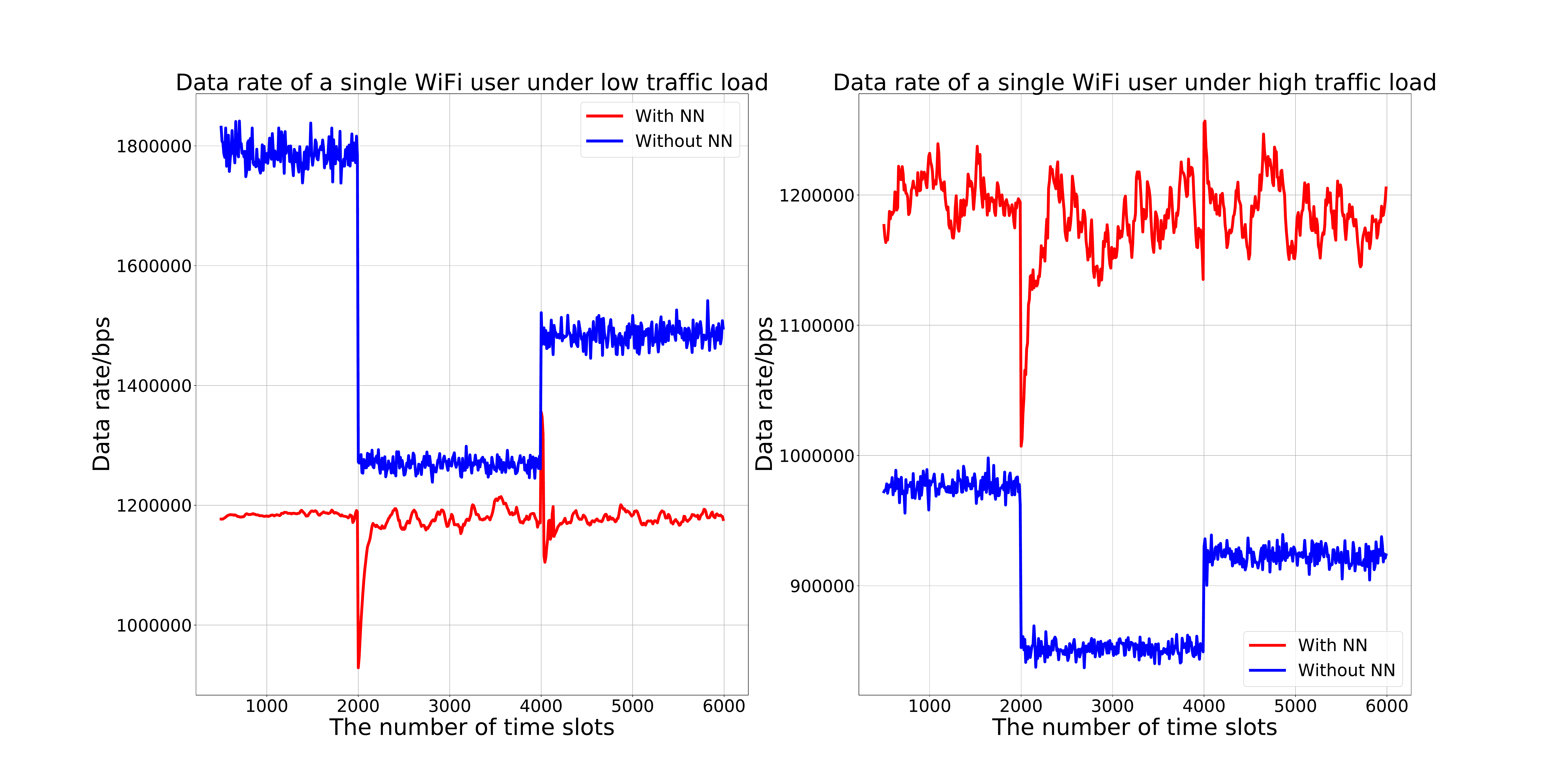}
\caption {Data rates of a single WiFi user.}\label{perDR}
\end{centering}
\end{figure}

Fig.~\ref{totalDR} and Fig.~\ref{perDR} are the simulation results of the total data rate on the channel and the data rate of a single WiFi user during the above detection period respectively.
The detailed simulation model can be found in \cite{19_Liu},
here NR-U devices access the unlicensed channel based on \emph{listen before talk} (LBT) mechanism. With the estimation result of NN, NR-U users can adjust the size of back-off window to use spectrum resources better while NR-U users without the information of traffic load are designed to use a fixed part of the unlicensed channel. Specifically, the fixed part is equal to half of the spectrum resources when the total data rate is maximum.
Simulation results in Fig.~\ref{totalDR} show that whether the WiFi traffic load is low or high, the proposed algorithm can maximize the total data rate on the channel. Moreover, as illustrated in Fig.~\ref{perDR}, without the known information, the WiFi users will lose performance significantly when the channel becomes busy, while with the NN prediction, the data rate of a single WiFi user can be kept basically the same. It means that when the channel is sensed idle, the NR-U users are able to compete for more spectrum resources on the basis of ensuring the quality of the WiFi transmission and when the WiFi traffic load is high, the NR-U users will reduce own transmission to achieve the harmonious coexistence with the WiFi system.

In terms of computational complexity, without a large number of loop computations in KF based method, the main computing task in Algorithm.~\ref{nn_rt} is to perform a forward propagation and a backward propagation on NN. Therefore, the computational complexity of the proposed approach decreases significantly than the KF based proposal. Table~\ref{time_compare} shows the time length of an update iteration and the average duration of a whole time slot on two different schemes when $n$ is over $20$.
As mentioned before, a whole time slot includes the time when NR-U user listens to the unlicensed spectrum for $K_a$ sub-frames and the time to execute the update iteration.
The sub-frames on the unlicensed bands can be described as
$T_s=192.58 \mu s$, $T_c=45.58\mu s$, $T_\delta=20\mu s$ and $K_{a}=100$, where $T_s$ is the time duration that the unlicensed channel is occupied by a successful transmission, $T_c$ denotes the period of a collided sub-frame on the channel and $T_{\sigma}$ is the duration of an empty sub-frame. Moreover, the update iterations of KF and NN are executed on Intel XEON CPU with $2.90GHz$ and NVIDIA Quadro P2000 GPU with calculate ability $6.1$, respectively.
Multiple results of operation show that the convergence speed of NN is much faster than the
extended KF based method, which verifies the decrease on the computational complexity of our proposal.

\begin{table}\centering
\small{
\caption{Running time comparison}\label{time_compare}
\vspace{0.1cm}
\begin{tabular}{|c|c|c|}
  \hline
  Simulation method & NN & KF\\
  \hline
  The duration of an update iteration & $7.42{\rm ms}$ & $67.16{\rm ms}$\\
  \hline
  The duration of a single time slot & $15.59{\rm ms}$ & $75.33{\rm ms}$\\
  \hline
\end{tabular}}

\end{table}

\section{Conclusion}\label{s5}
In this paper, an online unsupervised NN based approach to estimate the number of competing WiFi users on the unlicensed bands is proposed, where an NN is trained online as a filter to trade off the tested values against historical predictions. Compared with the conventional Kalman Filter scheme, the NN based method can achieve better accuracy and stable output when the system is congested. Besides, the computational complexity of the proposed method is much lower and the operation efficiency is improved significantly. Accurate estimation of the number of WiFi users in this work is of great significance to estimate WiFi traffic load, since in the NR-U system, cellular users can occupy the communication resources on the unlicensed channel adaptively according to the estimated WiFi traffic load without affecting the performance of original WiFi users.


\vspace{-1.4cm}

\profile[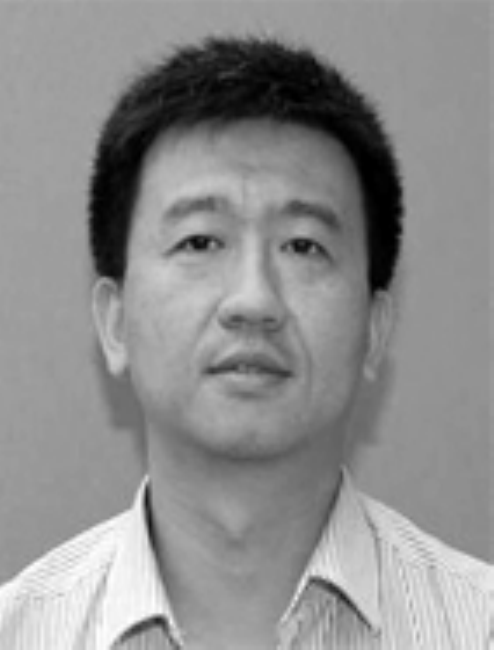]{Rin Yin}{
(M'13-S'19) received the B.S. degree in Computer Engineering from Yanbian University in 2001, China, the M.S. degree in Computer Engineering from KwaZulu-Natal University in Durban, South Africa in 2006, and the Ph.D. degree in Information and Electronic Engineering from Zhejiang University in 2011, respectively. From March 2011 to June 2013, he was a research fellow at the Department of Information and Electronic Engineering, Zhejiang University, China. He is now a professor in the School of Information and Electrical Engineering at Zhejiang University City College, China, and a joint honorary research fellow in the School of Electrical, Electronic and Computer Engineering at University of Kwa-Zulu Natal, South Africa. His research interests mainly focus on radio resource management in LTE unlicensed, millimeter wave cellular wireless networks, HetNet, cooperative communications, massive MIMO, optimization theory, game theory, and information theory. Prof. Yin regularly serves as the technical program committee (TPC) boards of prominent IEEE conferences such as ICC, GLOBECOM and PIMRC and chairs some of their technical sessions and reviwer for IEEE TWC, IEEE TVT, IEEE Tcom, IEEE Wireless Communications, IEEE Communications Magazine, IEEE Network, and IEEE TSP journals.}

\profile[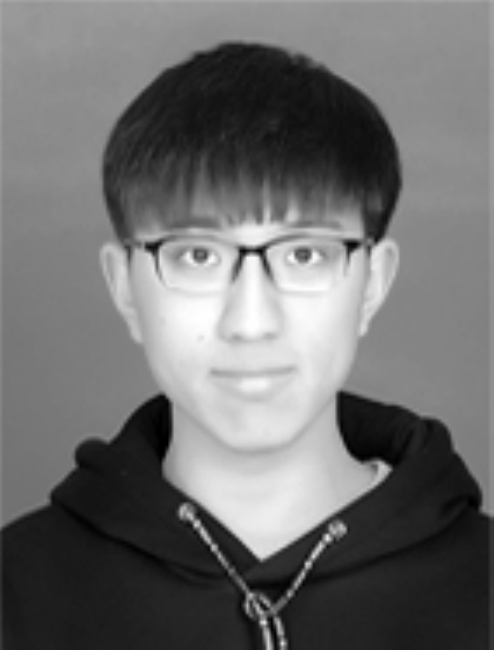]{Zhiqun Zou}{
received the B.E. from Zhejiang University, Hangzhou, China, in 2018. He is currently
pursuing the M.E. degree at the College of Information Science and Electronic Engineering (ISEE), Zhejiang University.
His research interests include D2D communications, Wireless communications and Machine learning applications.}

\profile[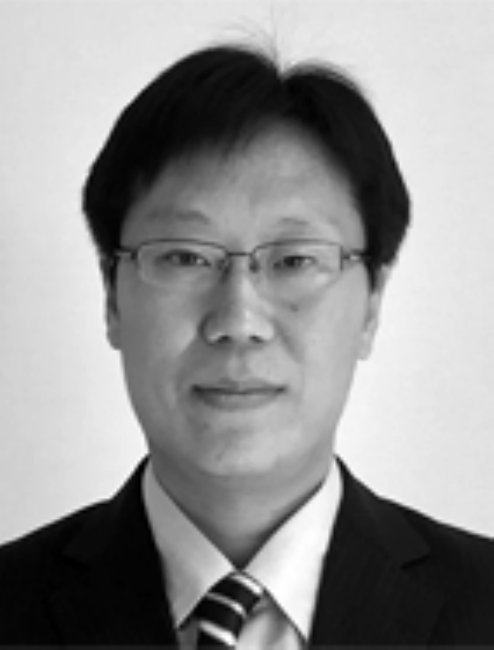]{Celimuge Wu}{
(Senior Member, IEEE) received the M.E. degree from the Beijing Institute of Technology, China, in 2006, and the Ph.D. degree
from The University of Electro-Communications, Japan, in 2010. He is currently an Associate
Professor with the Graduate School of Informatics and Engineering, The University of Electro-Communications.
His current research interests include vehicular networks, sensor networks, intelligent transport systems, the IoT, and mobile cloud computing. He is/has been the TPC Co-Chair of Wireless Days 2019,
ICT-DM 2019, and ICT-DM 2018. He is also the Chair of IEEE TCBD Special Interest Group on Big Data with Computational Intelligence. He is/hasbeen serving as an Associate Editor for IEEE ACCESS, the IEICE Transactions on Communications, the International Journal of Distributed Sensor Networks, and Sensors (MDPI), and a Guest Editor of the IEEE TRANSACTIONS
ON EMERGING TOPICS IN COMPUTATIONAL INTELLIGENCE, the IEEE Computational
Intelligence Magazine, and ACM/Springer MONET.}

\profile[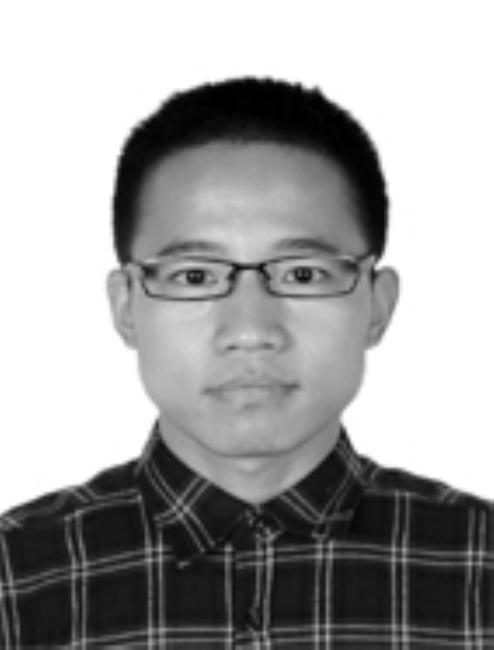]{Jiantao Yuan}{
received the B.E. degree in electronic and information engineering from Dalian University, Dalian, China, in 2009; the M.S. degree in signal and information processing from The First Research Institute of Telecommunications Technology, Shanghai, China, in 2012; and the
Ph.D degree in the college of information science and electrical engineer from Zhejiang University, Hangzhou, China.
He was with Datang mobile communication equipment co. LTD, Shanghai, China, from 2012 to 2013, where he was involved in LTE network planningand optimization.

He currently holds a post-doctoral position with the Institute of Ocean
Sensing and Networking of the Ocean College, Zhejiang University, Zhoushan, China. His research interests include cross-layer protocol design, 5G new-radio based access to unlicensed spectrum (NR-U) and ultra-reliable low latency communications.}

\profile[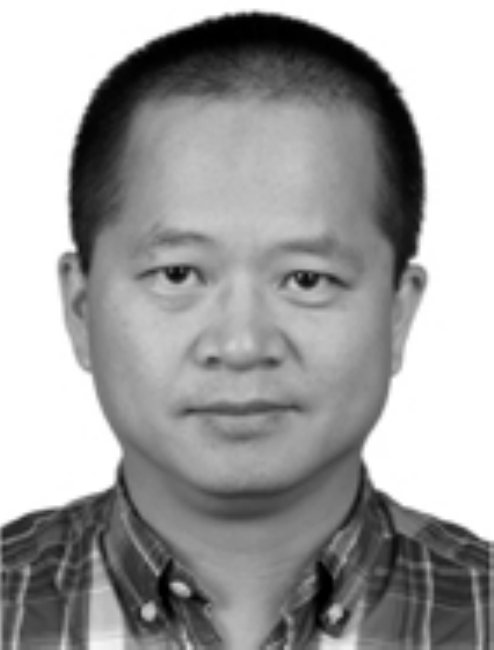]{Xianfu Chen}{
received his Ph.D. degree with honors in Signal and Information Processing, from the Department of Information Science and Electronic Engineering (ISEE) at Zhejiang University, Hangzhou, China, in March 2012.
Since April 2012, Dr. Chen has been with the VTT Technical Research Centre of Finland, Oulu, Finland, where he is currently a Senior Scientist.
His research interests cover various aspects of wireless communications and networking, with emphasis on human-level and artificial intelligence for resource awareness in next-generation communication networks.
Dr. Chen is serving and served as a Track Co-Chair and a TPC member for a number of IEEE ComSoc flagship conferences.
He is a Vice Chair of IEEE Special Interest Group on Big Data with Computational Intelligence, the members of which come from over 15 countries worldwide.
He is an IEEE member.}

\profile[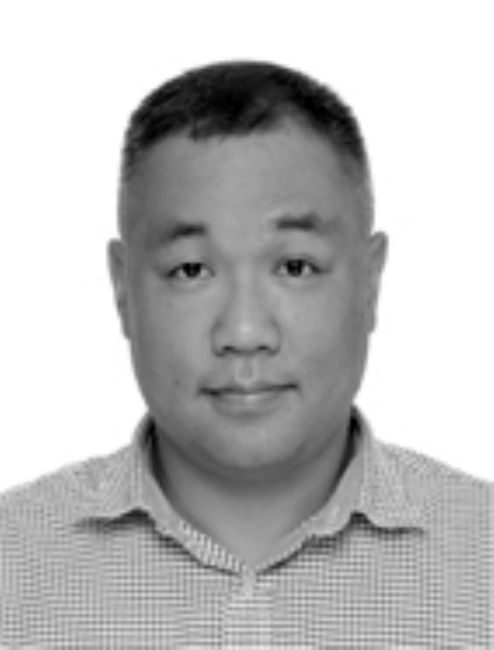]{Guanding Yu}{
(S'05-M'07-SM'13) received the B.E. and Ph.D. degrees in communication engineering from Zhejiang University, Hangzhou, China, in 2001 and 2006, respectively. He joined Zhejiang University in 2006, and is now a Full Professor with the College of Information and Electronic Engineering. From 2013 to 2015, he was also a Visiting Professor at the School of Electrical and Computer Engineering, Georgia Institute of Technology, Atlanta, GA, USA. His research interests include 5G communications and networks, mobile edge computing, and machine learning for wireless networks.

 Dr. Yu has served as a guest editor of \emph{IEEE Communications Magazine} special issue on Full-Duplex Communications, an editor of \emph{IEEE Journal on Selected Areas in Communications} Series on Green Communications and Networking, and a lead guest editor of \emph{IEEE Wireless Communications Magazine} special issue on LTE in Unlicensed Spectrum, and an Editor of IEEE Access. He is now serving as an editor of \emph{IEEE Transactions on Green Communications and Networking} and an editor of \emph{IEEE Wireless Communications Letters}. He received the 2016 IEEE ComSoc Asia-Pacific Outstanding Young Researcher Award. He regularly sits on the technical program committee (TPC) boards of prominent IEEE conferences such as ICC, GLOBECOM, and VTC. He also serves as a Symposium Co-Chair for IEEE Globecom 2019 and a Track Chair for IEEE VTC 2019'Fall.}

\end{document}